\documentclass[aps,pra,twocolumn,superscriptaddress,10pt,article,nofootinbib,showpacs]{revtex4-1}
\usepackage{amsfonts}
\usepackage{amsmath}
\usepackage{amsthm}
\usepackage{braket}
\usepackage{graphicx}
\usepackage{hyperref}
\usepackage{color}
\usepackage{amssymb}
\usepackage{dsfont}
\usepackage{verbatim}
\usepackage{amsmath,amscd}

\newcommand{\mathscr}[1]{\ensuremath{\mathcal{#1}}}

\newcommand{\bbra}[1]{\ensuremath{(#1|}}
\newcommand{\bket}[1]{\ensuremath{|#1)}}

\newcommand{\tr}[1]{\ensuremath{\text{Tr}[#1]}}

\newcommand{\hilb}{\ensuremath{\mathbb{H}}}

\newcommand{\mpo}{\ensuremath{{B}}}

\newcommand{\ZT}{\ensuremath{{\mathbb{Z}_2}}}

\newcommand{\sbra}[1]{\ensuremath{((#1|}}
\newcommand{\sket}[1]{\ensuremath{|#1))}}
\newcommand{\gr}[1]{{\ensuremath{|#1|}}}
\newcommand{\gra}{{\ensuremath{\mathfrak{g}}}}
\newcommand{\vect}[1]{\ensuremath{\underline{#1}}}
\newcommand{\N}{\ensuremath{{\mathcal{N}}}}
\newcommand{\V}{\ensuremath{{\mathbb{V}}}}
\newcommand{\subsp}{\ensuremath{{\mathcal{S}}}}

\newcommand{\pket}[1]{\ensuremath{|#1\rangle\rangle}}
\newcommand{\pbbra}[1]{\ensuremath{\bra{#1}}}
\newcommand{\pbket}[1]{\ensuremath{\ket{#1}}}

\begin{document}

\title{Fermionic Matrix Product Operators and Topological Phases of Matter}

\affiliation{Vienna Center for Quantum Technology, University of Vienna, Boltzmanngasse
5, 1090 Vienna, Austria}
\affiliation{Department of Physics and Astronomy, Ghent University, Krijgslaan 281
S9, B-9000 Ghent, Belgium}

\author{Dominic J. \surname{Williamson}}
\affiliation{Vienna Center for Quantum Technology, University of Vienna, Boltzmanngasse 5, 1090 Vienna, Austria}
\author{Nick \surname{Bultinck}}
\affiliation{Department of Physics and Astronomy, Ghent University, Krijgslaan 281 S9, B-9000 Ghent, Belgium}
\author{Jutho \surname{Haegeman}}
\affiliation{Department of Physics and Astronomy, Ghent University, Krijgslaan 281 S9, B-9000 Ghent, Belgium}
\author{Frank \surname{Verstraete}}
\affiliation{Vienna Center for Quantum Technology, University of Vienna, Boltzmanngasse 5, 1090 Vienna, Austria}
\affiliation{Department of Physics and Astronomy, Ghent University, Krijgslaan 281 S9, B-9000 Ghent, Belgium}

\begin{abstract}
We introduce the concept of fermionic matrix product operators, and show that they provide a natural representation of fermionic fusion tensor categories. This allows for the classification of two dimensional fermionic topological phases in  terms of matrix product operator algebras.
Using this approach we give a classification of fermionic symmetry protected topological phases with respect to a group $G$ in terms of three cohomology groups: $H^1(G,\mathbb{Z}_2)$, describing which matrix product operators are of Majorana type, $H^2(G,\mathbb{Z}_2)$, describing the fermionic nature of the fusion tensors that arise when two matrix product operators are multiplied, and the supercohomolgy group $\bar{H}^3(G,U(1))$ which corresponds to the associator that changes the order of fusion.
We also generalize the tensor network description of the string-net ground states to the fermionic setting, yielding simple representations of a class that includes the fermionic toric code.
\end{abstract}

\maketitle

\section{ Introduction}
\label{intro}

The search for topological phases of matter has recently empowered different fields of mathematical physics. From the technological point of view, such phases could be instrumental in creating fault tolerant architectures for quantum computation~\cite{kitaev2003fault,nayak2008non}. From the theoretical point of view, such phases are fascinating due to the fact that all the relevant physics is encoded in the entanglement structure of the corresponding many body wavefunctions~\cite{kitaev2006topological,levin2006detecting,chen2010local}. The natural field theoretic framework for classifying topological phases of matter is provided by topological quantum field theory.

For the bosonic case, explicit many-body realizations of non-chiral TQFTs have been constructed in terms of quantum doubles~\cite{kitaev2003fault} and string-nets~\cite{levin2005string}, whose ground state wavefunctions have a very simple realization in terms of projected entangled pair states~\cite{verstraete2006criticality,gu2009tensor,buerschaper2009explicit} which are a subclass of tensor network states~\cite{bridgeman2016hand}. This quantum tensor network representation has led to a mathematical framework in which the symmetries of the entanglement degrees of freedom are represented by an algebra of matrix product operators (MPO)~\cite{schuch2010peps,buerschaper2014twisted,csahinouglu2014characterizing,williamson2014matrix,bultinck2015anyons}. Those MPOs form  a representation of a fusion tensor category~\cite{bultinck2015anyons}, and the idempotents of a related MPO algebra yield an explicit construction of the different superselection sectors corresponding to the emergent anyons of the tensor network~\cite{bultinck2015anyons}. Similarly, bosonic symmetry protected topological (SPT) phases of matter have been characterized in terms of matrix product operators~\cite{chen2011two,williamson2014matrix}. Their classification can be understood in terms of the third cohomology group of the physical symmetry, which is an object that appears naturally when the fundamental theorem of MPOs is applied to a closed algebra of MPOs~\cite{cirac2016matrix}. One of the advantages of the MPO framework for describing topological phases is that it allows one to perturb away from the fixed point models, hence opening the possibility of studying topological phase transitions~\cite{haegeman2015shadows,marien2016condensation}.

For the case of fermionic topological quantum field theories, explicit realizations have also been obtained in terms of fermionic  string-nets~\cite{gu2014lattice,gu2015classification,ware2016ising,bhardwaj2016state} and fermionic SPTs~\cite{gu2014symmetry,cheng2015towards,tarantino2016discrete}. 
Yet a full understanding of the topological invariants and physical properties of fermionic phases remains to be developed. 

In this work, we introduce the concept of fermionic matrix product operators (fMPO) to describe the tensor network analogues of the fermionic string-nets and SPTs. The fermionic MPOs exhibit a natural $\ZT$-grading similar to the one obtained in the graded quantum inverse scattering method~\cite{kulish1980solutions,gohmann1998fermionic}. Furthermore the MPOs can be used to construct fermionic projected entangled pair states (fPEPS)~\cite{kraus2010fermionic,gu2010grassmann,pivzorn2010fermionic,wahl2013projected,corboz2010simulation,barthel2009contraction,zohar2016projected}  in a similar fashion to the bosonic case~\cite{bultinck2015anyons}. We find a classification of these fMPOs into discrete families.
In the case of symmetry-protected topological order this leads directly to a classification of physical phases of fermionic matter, some of which are not realizable in bosonic systems.

We note that the preliminary results provided her are superseded by our follow up work Ref.\cite{bultinck2017fermionic}.  

The manuscript is laid out as follows: in Section~\ref{fmpos} we review the basic structure of fermionic matrix product operators.
In Section~\ref{fusion} we focus on algebras of fermionic matrix product operators and the local fusion rules they imply.
In Section~\ref{spt} we specialize to the case of fermionic symmetry-protected topological order and find a classification of phases in terms of fermionic matrix product operators.
In Section~\ref{example} we show that the fermionic string nets are included in the fermionic matrix product operator injectivity formalism.

\section{Fermionic Matrix Product Operators}
\label{fmpos}

Matrix product operators (MPOs)~\cite{cirac2016matrix} constitute a natural class of operators with a 1D locality structure. This includes all 1D locality preserving unitaries which are known to be products of finite depth local unitary quantum circuits and translation operators~\cite{gross2012index}. Injective (or single blocked) MPOs form the irreducible constituents of a general MPO~\cite{fannes1992finitely,perez2007matrix}.
Injectivity can be rephrased as the requirement that the algebra generated by the local matrices is simple.
Two fundamental results underly the theory of MPOs, firstly after sufficient coarse graining any MPO can be written in a canonical form consisting of a sum of injective MPOs. Secondly if a pair of MPOs are equal for all system sizes there exits a gauge transformation relating the local tensors~\cite{fannes1992finitely,perez2007matrix,cirac2016matrix}.

To describe fermionic systems with MPOs we consider a $\ZT$-grading of the tensor indices, where $1\, (0)$ indicates the presence (absence) of a fermion.
For a homogeneous basis vector ${ \ket{i}\in \hilb_{\text{even}} \oplus \hilb_{\text{odd}} }$ we denote its parity by ${\gr{i}=0,1 }$. 
The $\ZT$ graded tensor product is denoted by $\ket{i}\otimes_\gra \ket{j}$. We use the following convention for larger tensor products
\begin{align}
\ket{i_1i_2\dots i_N} :&=\ket{i_1}\otimes_\gra\ket{i_2}\otimes_\gra \cdots\ket{i_N}
\\
\bra{j_1j_2\dots j_N}:&= \bra{j_1}\otimes_\gra\bra{j_2}\otimes_\gra\cdots\bra{j_N}
\end{align}
The inner product, or contraction of indices, is defined in the usual way
 \begin{align}
 \bra{i} \otimes_\gra \ket{j} &\mapsto \delta_{i,j}. 
 \end{align}
When multi-index tensors are contracted together, the Koszul sign rule for fermionic swaps 
\begin{align}
{ \ket{i}\otimes_\gra\ket{j} \leftrightarrow (-1)^{\gr{i} \gr{j}} \ket{j}\otimes_\gra \ket{i} },
\end{align}
 necessitates maintaining a careful ordering of the indices. We remark that this rule extends to arbitrary combinations of bras and kets. 
Hence one can derive another contraction rule
\begin{align}
  \ket{i} \otimes_\gra \bra{j} &\mapsto -\delta_{i,j}.
 \end{align}
 
 To ensure the global tensor network has a definite parity (without fine tuning) we consider only $\ZT$-even local tensors
\begin{equation}
\label{mpo}
\sum_{i j \alpha \beta} \mpo^{ij}_{\alpha \beta} \, \bket{\alpha}  \otimes_\gra \ket{i} \otimes_\gra \bra{j} \otimes_\gra \bbra{\beta }
\end{equation}
with $\gr{\alpha} + \gr{i} + \gr{j} + \gr{\beta} = 0 \mod 2$. Round kets indicate virtual indices of the MPO. Since the graded tensor product notation used above becomes overly cumbersome for products of many tensors, we will henceforth neglect to explicitly write $\otimes_\gra$ in most instances. For example, the expression $ \bket{\alpha}  \otimes_\gra \ket{i} \otimes_\gra \bra{j} \otimes_\gra \bbra{\beta }$  becomes $ \bket{\alpha}  \ket{i} \bra{j} \bbra{\beta }$.

The tensor in Eq.\eqref{mpo} can be used to build a MPO of length $L$ by taking the graded tensor product of $L$ copies and contracting consecutive virtual indices $\bbra{\beta_{i-1}}\otimes_\gra \bket{\alpha_{i}}\mapsto \delta_{\beta_{i-1},\alpha_{i}}$. This results in an MPO with two remaining virtual indices at its boundaries 
\begin{align*}
\sum_{\vect{i}\, \vect{j} \alpha_1 \beta_L } [{\mpo^{i_1j_1} \cdots \mpo^{i_L j_L}}]_{\alpha_1\beta_L} \, \bket{\alpha_1} \ket{i_1} \bra{j_1} \otimes_\gra \cdots \ket{i_L} \bra{j_L} \bbra{\beta_L }.
\end{align*}

Closing an MPO with periodic boundary conditions gives rise to a translation invariant even fermion operator. On the virtual level, this results in a supertrace, generalizing the usual trace used in the bosonic case
\begin{align}
%\sum_{\vect{i}\, \vect{j} \alpha } [{\mpo^{i_1j_1} \cdots \mpo^{i_L j_L}}]_{\alpha\alpha} \, \bket{\alpha} \ket{i_1} \bra{j_1} \otimes_\gra \cdots \ket{i_L} \bra{j_L} \bbra{\alpha }
%\nonumber \\
&\sum_{\vect{i}\, \vect{j} \alpha \beta} (-1)^{\gr{\beta}}[{\mpo^{i_1j_1} \cdots \mpo^{i_L j_L}}]_{\alpha\beta}   \bbra{\beta }\bket{\alpha} \ket{i_1} \bra{j_1} \otimes_\gra \cdots \ket{i_L} \bra{j_L}
\nonumber \\
&\mapsto
\sum_{\vect{i}\, \vect{j} } \tr{\Lambda \mpo^{i_1j_1} \cdots \mpo^{i_L j_L}} \, \ket{i_1} \bra{j_1} \otimes_\gra \cdots \ket{i_L} \bra{j_L},
\end{align}
where $\Lambda \ket{a} = (-1)^{\gr{a}} \ket{a}$.

The inclusion of fermionic modes means the algebras defining irreducible MPOs are $\ZT$-graded simple algebras, which can be of odd or even type~\cite{fidkowski2011topological,Wall}:
\begin{equation}
\begin{tabular}{ c | c | c  | c}
 & Even element & Odd element & Center
\\ \hline
Even & $\left(\begin{matrix}
A^{ij} & 0
\\
0 & B^{ij}
\end{matrix}\right)$ & $\left(\begin{matrix}
0 & C^{ij}
\\
D^{ij} & 0
\end{matrix}\right)$ & $\openone$
\\ \hline
Odd & $\left(\begin{matrix}
A^{ij} & 0
\\
0 & A^{ij}
\end{matrix}\right)$ & $\left(\begin{matrix}
0 & B^{ij}
\\
-B^{ij} & 0
\end{matrix}\right)$ & $\openone$, $Y = \left(\begin{matrix} 0 & \openone \\ -\openone & 0\end{matrix}\right)$
\end{tabular}.
\end{equation}	
See Refs.~\cite{BultinckPrep,kapustin2016spin} for a detailed treatment of irreducible fermionic matrix product states, which carries over to fMPOs. The odd algebras correspond to the phenomena of Majorana edge modes~\cite{kitaev2001unpaired} and we henceforth refer to an irreducible odd algebra fMPO as a Majorana fMPO. Such MPOs are not irreducible in the setting of bosonic spin systems, since they have a non-trivial center. In the bosonic case they would correspond to ``cat states'' which consist of a macroscopic superposition of locally distinguishable configurations. However, as all physical local operators obey a superselection rule in the fermionic case, they cannot distinguish the two macroscopic configurations~\cite{verstraete2003quantum,van2015entanglement}.

A Majorana fMPO yields zero when closed with the supertrace, but adding the non-trivial center $Y$ in the fMPO definition still gives rise to a non-zero TI operator which now has odd fermion parity. Vice versa, the even algebra fMPO cannot yield an odd TI operator. With anti-periodic boundary conditions instead, both types of irreducible fMPO give rise to an even operator \cite{BultinckPrep,kapustin2016spin}. To develop a full theory of virtual fMPO symmetries of an fPEPS including both types of boundary conditions is extremely important~\cite{bultinck2017fermionic}. However, in this preliminary work we focus our attention on fMPOs with periodic boundary conditions.

\section{Algebra of \lowercase{f}MPOs and fusion}
\label{fusion}

In this section we study algebras of fMPOs. We outline how the algebra structure leads to local fusion tensors. Different fusion paths are related by an associator that satisfies the fermionic pentagon equation. Hence we find a classification of fMPO algebras in terms of equivalence classes of solutions to the fermionic pentagon equation.

%Following the approach of Ref.\cite{bultinck2015anyons} we consider projection fMPOs that can be decomposed into irreducible constituents
%\begin{equation}
% \mpo^{ij} =\bigoplus_{a=1}^{\N} \mpo^{ij}_{a}
%\end{equation}
%where each set of matrices $\mpo^{ij}_{a}$ spans a simple graded algebra.
We introduce a family of irreducible TI fMPOs, labeled $a$, such that $O_a^L$ acts on a chain of length $L$ as
\begin{equation*}
O_a^L=\sum_{\vect{i}\, \vect{j} } \tr{\Delta_a \mpo_a^{i_1j_1} \cdots \mpo_a^{i_L j_L}} \, \ket{i_1} \bra{j_1} \otimes_\gra \cdots \ket{i_L} \bra{j_L}
.
\end{equation*}
where $\Delta_a=Y$ or $\Delta_a=\Lambda$ depending on $a$ being a Majorana or non-Majorana fMPO. This family forms a closed algebra if
\begin{equation}
O_a^L O_b^L = \sum_{c=1}^{\N} N_{ab}^c O_c^L
\end{equation}
where $N_{ab}^c$ have to be integers because they result from the canonical decomposition of the left hand side. Furthermore, in the case of TI fMPOs, the fMPO algebra has an additional $\ZT$-graded structure. The product of two TI Majorana fMPOs is non Majorana, and the product of a Majorana and non Majorana TI fMPO is again Majorana, because of the one-to-one correspondence between fMPO type and the global fermion parity of the TI operator $O_a^L$. We remark that this no longer holds in the case of antiperiodic boundary conditions\cite{bultinck2017fermionic,aasen2017fermion}. 

The multiplication rules obeyed by the irreducible fMPOs imply the existence of $\ZT$-even local fusion tensors 
\begin{align}
{ X_{ab}^{c}:\V_{ab}^c \otimes_\gra \hilb_c\rightarrow \hilb_a \otimes_\gra \hilb_b }
\end{align}
 that relate products of individual fMPO tensors. Here $\V_{ab}^c$ is a $\ZT$-graded Hilbert space with $N_{ab}^c$ basis elements given by degeneracy labels $\{\mu\}$ and $\hilb_a$ denotes the virtual Hilbert space of the fMPO $a$.

When considering products of fMPOs it is convenient to work with a different choice of local basis for each individual fMPO tensor as follows
\begin{equation}
\label{mpob}
\sum_{i j \alpha \beta} [\mpo_a]^{ij}_{\alpha \beta} \,  \ket{i} \bket{\alpha} \bra{j} \bbra{\beta }
\end{equation}
we continue to use this choice for the remainder of the section. Note the tensor in Eq.\eqref{mpob} is equal to the tensor with numerical coefficient $(-1)^{\gr{i}\gr{\alpha}}[\mpo_a]^{ij}_{\alpha \beta}$ for the basis used in Eq.\eqref{mpo}.
Hence the local tensor for $O_a^L O_b^L$ is
\begin{align}
\sum_{\substack{i j k \\\alpha \beta \gamma \delta}} [\mpo_a]^{ij}_{\alpha \beta} \,  \ket{i} \bket{\alpha} \bra{j} \bbra{\beta } \otimes_\gra  [\mpo_b]^{jk}_{\gamma \delta}  \ket{j} \bket{\gamma} \bra{k} \bbra{\delta }
\nonumber \\
= \sum_{\substack{i k \\\alpha \beta \gamma \delta}} \sum_j [\mpo_a]^{ij}_{\alpha \beta} [\mpo_b]^{jk}_{\gamma \delta} \, \ket{i} \bket{\alpha \gamma} \bra{k} \bbra{\delta \beta  }
\end{align}
which is related to the local tensor for $O^L_c$ by a gauge transformation as follows
\begin{align}
& \sum_{\substack{i k \alpha \beta \\ \gamma \delta \sigma \rho }}  [({X_{ab}^{c}})^{-1}]_{\gamma \alpha}^{\sigma,\mu} \sket{\mu}\bket{\sigma} \bbra{\gamma \alpha} \otimes_\gra
 \sum_j [\mpo_a]^{ij}_{\alpha \beta} [\mpo_b]^{jk}_{\gamma \delta} \,
  \nonumber \\
& \phantom{\sum\sum\sum} \ket{i} \bket{\alpha \gamma} \bra{k} \bbra{\delta \beta  }  \otimes_\gra
  [X_{ab}^{c}]_{\beta \delta}^{\rho, \mu} \bket{\beta \delta} \bbra{\rho} \sbra{\mu}
  \nonumber \\
&  =
 \sum_{{i k \sigma \rho }}  \sum_{j \alpha \beta \gamma \delta} (-1)^{\gr{\mu}\gr{i}} [{(X_{ab}^{c})}^{-1}]_{\gamma \alpha}^{\sigma,\mu}
 [\mpo_a]^{ij}_{\alpha \beta} [\mpo_b]^{jk}_{\gamma \delta}
  [X_{ab}^{c}]_{\beta \delta}^{\rho, \mu}
  \nonumber \\
&\phantom{\sum\sum\sum\sum\sum\sum\sum} \ket{i} \bket{\sigma} \bra{k} \bbra{\rho  }
 \otimes_\gra \sket{\mu} \sbra{\mu}
  \nonumber \\
 & = \sum_{i k \sigma \rho} [\mpo_c]^{ik}_{\sigma \rho} \,  \ket{i} \bket{\sigma} \bra{k} \bbra{\rho }  \otimes_\gra \sket{\mu} \sbra{\mu}.
\end{align}
The double rounded kets denote degeneracy indices in the above equations. 
This furthermore implies
\begin{align}
\label{reduction}
\bigoplus_{\mu=1}^{N_{ab}^c} (-1)^{\gr{\mu}\gr{i}} [{(X_{ab}^{c})}^{-1}]^{\mu}
 \sum_{j }  [\mpo_a]^{ij} \otimes [\mpo_b]^{jk} \,
  [X_{ab}^{c}]^{\mu}
  \nonumber \\
   = \openone_{N_{ab}^{c}} \otimes [\mpo_c]^{ik},
\end{align}
where $ [\mpo_a]^{ij}$ is a bosonic matrix with coefficients $ [\mpo_a]^{ij}_{\alpha\beta}$. 
Hence the fusion tensors $X_{ab}^{c}$ are only defined up to multiplication by an invertible $\ZT$-even matrix on the degeneracy index $\mu$. 
For Majorana fMPOs the analysis is complicated considerably by additional relations between fusion tensors induced by pushing a $Y$ matrix onto the tensor from any of the Majorana fMPO internal indices. A full treatment of this is beyond the scope of the current work and is instead given in Ref.\cite{bultinck2017fermionic}. 

At this point we make the simplifying assumption that the fusion tensors are invertible on the subspace relevant to the fMPOs $\sum\limits_c{X_{ab}^c (X_{ab}^{c})^{-1} = \openone_{\subsp_{a\times b}}}$. Where $\openone_{\subsp_{a\times b}}$ is the projector onto the support subspace of the fMPO virtual indices, i.e. ${\openone_{\subsp_{a\times b}} \sum_{j }  \mpo_a^{ij} \otimes_\gra \mpo_b^{jk}= \sum_{j }  \mpo_a^{ij} \otimes_\gra \mpo_b^{jk}}$.
This corresponds to the assumption that no off diagonal blocks, which couple different irreducible fMPOs, appear in the product of fMPO tensors. Note such off diagonal blocks do not contribute to the fMPO once the boundary conditions are closed.

Now we have a stronger version of Eq.\eqref{reduction} known as the \emph{zipper condition}~\cite{williamson2014matrix,bultinck2015anyons}
\begin{align}
\label{zipper}
 \sum_{j }  [\mpo_a]^{ij} \otimes [\mpo_b]^{jk} \,
  [X_{ab}^{c}]^{\mu}
   =  (-1)^{\gr{\mu}\gr{i}}   [X_{ab}^{c}]^{\mu}   [\mpo_c]^{ik}
\end{align}
equivalently
\begin{align}
\label{zippertwo}
 \sum_{j }  [\mpo_a]^{ij} \otimes [\mpo_b]^{jk} =   \sum_\mu (-1)^{\gr{\mu}\gr{i}} [X_{ab}^{c}]^{\mu}  [\mpo_c]^{ik}  [{(X_{ab}^{c})}^{-1}]^{\mu}
\end{align}
which corresponds to splitting an fMPO $O^L_c$ into $O^L_aO^L_b$.

The product of three fMPOs $O^L_a,O^L_b,O^L_c$ may be fused in two different ways, $(a\times b)\times c$ or $a \times (b\times c)$. If the outcome of the product is an injective fMPO $O^L_d$ we find that the products of fusion tensors ${\bigoplus_{f} X^f_{ab} X^d_{fc}}$ and ${\bigoplus_{e} X^e_{bc}X^d_{ae}}$ are related by a $\ZT$-even tensor, known as the associator or $F$-symbol, acting on the degeneracy indices.
To find the $F$-symbol we consider the equality between the two splittings of $B_d$, $(B_a \otimes_\gra B_b) \otimes_\gra B_c$ and $B_a \otimes_\gra (B_b \otimes_\gra B_c)$ respectively. 
At this, point to simplify the analysis considerably, we assume that the fMPOs $O^L_d$ is not of Majorana type. With this assumption we can apply the injective fMPO tensor's inverse
\begin{equation}
\sum_{ij\alpha\beta} (-1)^{\gr{i}\gr{\beta}+\gr{i}+\gr{\beta}} [B_d^{-1}]^{ij}_{\alpha\beta} \bket{\beta} \ket{j} \bbra{\alpha} \bra{i}
\end{equation}
 which satisfies ${\sum_{ij} (-1)^{\gr{i}\gr{\beta}}[B_d]^{ij}_{\alpha \beta}[B_d^{-1}]^{ij}_{\alpha' \beta'}=\delta_{\alpha\alpha'}\delta_{\beta\beta'}}$, to remove $B_d$ from the equation. For a full analysis of the general case see Ref.\cite{bultinck2017fermionic}. This leaves us with
\begin{align}
\sum_{f \theta\theta' \mu \nu}
[X_{ab}^f]^{\theta,\mu}_{\alpha \beta} [X_{fc}^d]^{\delta,\nu}_{\theta \gamma}
[(X_{fc}^d)^{-1}]^{\delta',\nu}_{\theta' \gamma'} [(X_{ab}^f)^{-1}]^{\theta' ,\mu}_{\alpha' \beta'}
\nonumber \\
 = \sum_{e \psi \psi' \sigma \rho}  (-1)^{\gr{\sigma}(\gr{\alpha}+\gr{\alpha'})}
[X_{bc}^e]^{\psi,\sigma}_{\beta \gamma} [X_{ae}^d]^{\delta,\rho}_{\alpha \psi}
\nonumber \\
[(X_{ae}^d)^{-1}]^{\delta',\rho}_{\alpha' \psi'} [(X_{bc}^e)^{-1}]^{\psi',\sigma}_{\beta' \gamma'}
\end{align}
multiplying with ${\sum\limits_{e' \psi'' \sigma' \rho'}  (-1)^{\gr{\sigma'}\gr{\alpha'}}
[X_{bc}^{e'}]^{\psi'',\sigma'}_{\beta' \gamma'} [X_{ae'}^{d'}]^{\delta'',\rho'}_{\alpha' \psi''} }$ from the right (and summing over $\alpha',\beta',\gamma'$) results in several delta conditions $\delta_{\sigma\sigma'}\delta_{\rho\rho'}\delta_{dd'}\delta_{ee'}\delta_{\psi'\psi''}\delta_{\delta'\delta''}$.
We then trace out the $\delta'$ degree of freedom to find
\begin{align}
&\sum_{f  \mu \nu \theta}
[X_{ab}^f]^{\theta,\mu}_{\alpha \beta} [X_{fc}^d]^{\delta,\nu}_{\theta \gamma}
[F^{cba}_{d}]_{e\sigma \rho}^{f \mu \nu}
\nonumber \\
 &\phantom{\sum} = \sum_{\psi }  (-1)^{\gr{\sigma}\gr{\alpha}}
[X_{bc}^e]^{\psi,\sigma}_{\beta \gamma} [X_{ae}^d]^{\delta,\rho}_{\alpha \psi}
\end{align}
where the $F$-symbol is given by
\begin{align}
[F^{cba}_{d}]_{e\sigma \rho}^{f \mu \nu}
:= \sum_{\substack{\alpha'\beta'\gamma'\delta'\theta'\psi'}}
\frac{ (-1)^{\gr{\sigma}\gr{\alpha'}}}{\tr{\openone_d}}
[(X_{fc}^d)^{-1}]^{\delta',\nu}_{\theta' \gamma'}
\nonumber \\
 [(X_{ab}^f)^{-1}]^{\theta' ,\mu}_{\alpha' \beta'}  [X_{bc}^{e}]^{\psi',\sigma}_{\beta' \gamma'} [X_{ae}^d]^{\delta',\rho}_{\alpha' \psi'} .
\end{align}
This associator is only defined up to a gauge transformation consisting of an invertible $\ZT$-even matrix on each of the degeneracy indices.

Considering the different ways in which a product of four fMPOs can be reduced leads to a consistency constraint on the $F$-symbols known as the fermionic pentagon equation. Since $((a\times b)\times c)\times d$ can be transformed to $a\times (b\times(c\times d))$ via either $(a\times b) \times (c\times d)$ or $(a \times (b\times c))\times d$ then $a\times ((b\times c)\times d)$ we find the consistency equation
\begin{align}
\label{fpentagon}
&\sum_\delta (-1)^{\gr{\nu}\gr{\alpha}} [F^{fcd}_{e}]_{g\beta \gamma}^{l \nu \delta} [F^{abl}_{e}]_{f\alpha \delta}^{k \lambda \mu}
\nonumber \\
& \phantom{\sum} =  \sum_{h \sigma \psi \rho} [F^{abc}_{g}]_{f\alpha \beta}^{h \sigma \psi} [F^{ahd}_{e}]_{g\psi\gamma}^{k\rho\mu} [F^{bcd}_{k}]_{h \sigma \rho}^{l \nu \lambda}
\end{align}
which precisely matches the fermionic pentagon equation~\cite{gu2014lattice,gu2015classification,lan2015theory}. A result known as Mac Lane's coherence theorem implies the pentagon equation constraint alone is sufficient to ensure all other possible consistency relations are satisfied~\cite{mac2013categories}. 
Furthermore solutions of the fermionic pentagon equation, up to gauge equivalence, fall into  discrete families, a property known as Ocneanu rigidity~\cite{kitaev2006anyons,mac2013categories,etingof2005fusion}.

At this point we have completed our study of fMPO algebras satisfying the \emph{zipper condition}. In the following sections we give two applications in the context of topological phases of matter. We start with symmetry-protected topological phases and proceed to fermionic string-nets.

\section{\lowercase{f}MPO group representations and classification of 2D SPT phases}
\label{spt}

For two-dimensional fermionic symmetry-protected topological (SPT) phases, following Ref.\cite{williamson2014matrix}, the relevant problem is to classify inequivalent fMPO representations of the discrete, unitary, physical symmetry group~$G$. These correspond to possible anomalous symmetry actions on the boundary of the SPT phase.
The constraints of the general classification as presented in sections \ref{fmpos} and \ref{fusion} reduce to a classification in terms of group (super) cohomology and a group homomorphism.
In this section we derive the classification and also describe a tensor network interpretation of each label in the classification.

The starting point is a collection of fMPOs $\{O^L_g\}$ of length $L$, where $g\in G$. As explained in section \ref{fmpos} each MPO may be of Majorana type or not which leads to a $\ZT$-grading of the group multiplication. This is described by a 1-cocycle in $H^{1}(G,\ZT)$.
Since the MPO group representation satisfies $O^L_gO^L_h=O^L_{gh}$ the fusion is unique and has no degeneracy. At this point we specialise to the case where none of the fMPOs in the representation are of Majorana type, for a complete treatment see Ref.\cite{bultinck2017fermionic}. 
Hence the fusion tensor takes the form
\begin{equation}
\sum_{\alpha \beta \gamma}  [X_{g,h}]_{\alpha \beta }^{\gamma} \bket{\alpha \beta} \bbra{\gamma} \sbra{Z(g,h)}
\end{equation}
where $Z(g,h)=\gr{\alpha}+\gr{\beta}+\gr{\gamma}$ is a $\ZT$ valued function on $G\times G$ describing which fusions are fermionic and which are bosonic.

The $F$-symbol now acts on a one dimensional space and is hence a complex phase
\begin{align}
&[F^{k,h,g}_{ghk}]_{hk,Z(h,k)Z(g,hk)}^{gh,Z(g,h)Z(gh,k)} \sket{Z(h,k)Z(g,hk)}\sbra{Z(g,h)Z(gh,k)}
\nonumber \\
&\phantom{\sum}= \alpha(g,h,k) \sket{Z(h,k)Z(g,hk)}\sbra{Z(g,h)Z(gh,k)}
\end{align}
The evenness constraint on the $F$-symbol associator becomes a 2-cocycle constraint on the fusion parity function
\begin{equation}
Z(h,k)+Z(g,hk)=Z(g,h)+Z(gh,k)
\end{equation}
 Since the fusion parity is only defined up to shifting the parity of the virtual indices of individual MPO tensors the relevant equivalence classes are elements of $H^2(G,\ZT)$. We remark that if Majorana type fMPOs are included this $H^2$ label is only well defined for the trivial (non Majorana) subgroup~\cite{bultinck2017fermionic}. 

The pentagon equation then reduces to the super {3-cocycle} condition for $\alpha$~\cite{gu2014symmetry}
\begin{equation}
\label{supercoho}
\frac{\alpha(g,h,k)\alpha(g,hk,l)\alpha(h,k,l)}{\alpha(gh,k,l)\alpha(g,h,kl)}=(-1)^{Z(g,h)Z(k,l)}.
\end{equation}
Since $\alpha$ is only defined up to a rephasing of the fusion tensors the relevant equivalence classes are elements of the supercohomology group $\bar H^3(G,U(1))$. For a general definition of the $\bar H^3$ label for group representations involving Majorana fMPOs we defer to Ref.\cite{bultinck2017fermionic}. 

We remark that Eq.\eqref{supercoho} is an inhomogeneous version of the regular 3-cocycle equation, and the product (or ratio) of any pair of solutions to Eq.\eqref{supercoho} is itself a normal 3-cocycle. The supercocycles are readily calculated using elementary algebra and their equivalence classes may differ from normal cocycles. For example
\begin{equation}
\begin{tabular}{ c | c | c | c  }
$G$ & $H^2(G,\ZT)$ & $H^3(G,U(1))$ & $\bar H^3(G,U(1))$
\\ \hline
$\ZT$ & $\ZT$ & $\ZT$ & $\mathbb{Z}_4$
\\ \hline
$\ZT \times \ZT$ & $\mathbb{Z}_2^3$ & $\mathbb{Z}_2^3$ & $ \mathbb{Z}_4^3$
\\ \hline
$A_4$ & $\ZT$ &  $\mathbb{Z}_6$ &  $\mathbb{Z}_{12}$
\end{tabular}.
\end{equation}

Hence we have derived the classification of (2+1)D SPT in terms of an element in $H^1(G,\ZT)$ that describes which MPOs are Majorana, an element in $H^2(G,\ZT)$ describing the fermionic nature of the fusion tensors and an element of $\bar H^3(G,U(1))$ describing the associator which changes the order of fusion~\cite{gu2014symmetry,cheng2015towards}. Note there is an additional constraint on the relevant elements $Z\in H^2(G,\ZT)$, namely that there exists an $\alpha$ satisfying Eq.\eqref{supercoho} for that choice of $Z$.

An advantage of the fMPO approach is that it allows us to easily calculate the group structure of the (2+1)D fermionic SPT phases under stacking.
Recall the bosonic stacking relation is simply pointwise multiplication of cocycles, the result of stacking two fermionic SPTs ${(\alpha'',Z''):=(\alpha,Z)\otimes(\alpha',Z')}$ has fusion tensors given by
\begin{align}
&\sum_{{\alpha \beta \gamma  \alpha' \beta' \gamma'}}  [X_{g,h}]_{\alpha \beta }^{\gamma} \bket{\alpha \beta} \bbra{\gamma} \sbra{Z(g,h)}  
\nonumber \\
& \phantom{\sum\sum\sum}
 \otimes  [X'_{g,h}]_{\alpha' \beta' }^{\gamma'} \bket{\alpha' \beta'} \bbra{\gamma'} \sbra{Z'(g,h)} \\
\cong & \sum_{{\alpha \beta \gamma  \alpha' \beta' \gamma'}}  (-1)^{\gr{\alpha'}\gr{\beta}+\gr{\gamma}Z'(g,h)} [X_{g,h}]_{\alpha \beta }^{\gamma}   [X'_{g,h}]_{\alpha' \beta' }^{\gamma'}  
\nonumber \\
& \phantom{\sum\sum\sum}
\bket{\alpha \alpha' \beta \beta'}\bbra{\gamma' \gamma} \sbra{Z(g,h)+Z'(g,h)}
\end{align} 
When considering the reduction of $(gh)k$, some rearrangement of fermion modes is required to bring together $X_{g,h}X_{gh,k}$ and $X'_{g,h}X'_{gh,k}$ one may then apply the individual associators before further rearrangement to match the reduction $g(hk)$. This leads to the formula
\begin{align}
\alpha''(g,h,k)&=(-1)^{Z(gh,k)Z'(g,h)+Z(g,hk)Z'(h,k)}
\nonumber \\
&\phantom{==}\alpha(g,h,k)\alpha'(g,h,k) \label{composition}
\\
Z''(g,h)&=Z(g,h)+Z'(g,h) \label{composition2}
\end{align}
for the stacked SPT phase, which matches the result of Refs.\cite{lan2015theory,bhardwaj2016state}. We remark that this group of Gu-Wen type fermionic SPT phases under stacking is a possibly nontrivial $H^2(G,\mathbb{Z}_2)$-extension of the group $H^3(G,U(1))$ of bosonic SPT phases under stacking. 
For a non-trivial $H^1(G,\mathbb{Z}_2)$ label the homomorphism $f'':G\rightarrow \mathbb{Z}_2$ of the composite fMPO is simply the sum of the homomorphisms $f$ and $f'$ of the stacked fMPOs, i.e. $f'' = f + f'$.
Moreover, this may lead to a further nontrivial $\mathbb{Z}_2$-extension of the group $\bar H^3(G,U(1))$ of Gu-Wen type fermionic phases under stacking, see Ref.\cite{bultinck2017fermionic} for further explanation. This already occurs for $\mathbb{Z}_2$ fermionic SPTs, for which we have ${ H^3(\mathbb{Z}_2,U(1))=\mathbb{Z}_2},$ ${\bar H^3(\mathbb{Z}_2,U(1))=\mathbb{Z}_4},$ and the full classification given by~$\mathbb{Z}_8$. For a tensor network description of the root phase of this classification see Refs.\cite{bultinck2017fermionic,aasen2017fermion}.

We close the section by discussing physical interpretations of each label appearing in the classification of single block fMPOs group representations. Recall the physical interpretation of a nontrivial $H^3$ label for a bosonic SPT is the appearance of symmetry-protected gapless edge modes~\cite{levin2012braiding}. In PEPS this corresponds to the fact that there cannot be a single block MPS fixed point under an MPO group action with nontrivial $H^3$ label~\cite{chen2011two}. For fermionic SPT phases this interpretation of the $\bar H^3$ label carries over directly.

The $H^2$ label also has a physical interpretation as the projective representation of the symmetry carried by the edge of a cylinder with a fermion parity flux insertion (symmetry twist). This appears in the fMPO framework since a set of group fMPOs form a projective representation, with factor set $(-1)^{Z(g,h)}$, when closed with the supertrace ($\Lambda$) on periodic boundary conditions.

Finally, $H^1$ can be interpreted as specifying which symmetry defects are non-Abelian due to the localization of an odd number of Majorana modes, it is also related to the appearance of Majorana edge modes~\cite{cheng2015towards}.
In the fMPO framework, a nontrivial $H^1$ index implies there is no single block fMPS fixed point under the full fMPO group action. 
To see this, first notice that a nontrivial $H^1$ label implies that there is a Majorana fMPO representation of $\ZT$ within the full MPO group action. Since acting with an odd operator flips an even state to odd and vice versa there can be no single block MPS that is symmetric under such a Majorana fMPO. 
Furthermore, an fPEPS state on a cylinder that has been twisted by fMPOs in the Majorana sector supports Majorana edge modes, see Ref.\cite{bultinck2017fermionic}.

\section{Example: Fermionic string-nets}
\label{example}

In this section we demonstrate that fMPO algebras also appear in the context of fermionic string-net models~\cite{levin2005string,gu2014lattice,gu2015classification,bhardwaj2016state}. We furthermore show that the relevant fMPOs satisfy the zipper condition, and the resulting associator matches the $F$-symbol used as input for the construction.

The fermionic $F$-symbol arises as a basis transformation matrix between fusion trees of a fermionic fusion category~\cite{gu2014lattice,gu2015classification,lan2015theory}
\begin{equation}
\vcenter{\hbox{
\includegraphics[scale=.95]{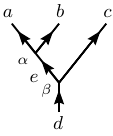}}} \bra{\beta \alpha} = \sum_{f \mu \nu} [F_{d}^{abc}]_{e\alpha\beta}^{f\mu\nu} \vcenter{\hbox{
\includegraphics[scale=.95]{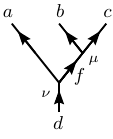}}} \bra{\nu \mu}
\end{equation}
where we have explicitly specified an ordering of the splitting degeneracy degrees of freedom to emphasize their fermionic nature.
The $F$-symbol is a nonsingular matrix on the fusion spaces with inverse
\begin{equation}
\label{finv}
 \sum\limits_{{e \alpha \beta , f \mu \nu}} [(F_{d}^{abc})^{-1}]^{e\alpha\beta}_{f\mu\nu} \ket{f;\mu \nu}\bra{e;\beta\alpha}
\end{equation}
note $e$ and $f$ are bosonic modes. For the fusion theory to be unitary the $F$-symbol must satisfy
$$[(F_{d}^{abc})^{-1}]^{e\alpha\beta}_{f\mu\nu}=([F_{d}^{abc}]_{e\alpha\beta}^{f\mu\nu})^* . $$
Furthermore for the $F$-symbol to define a consistent fermionic fusion category it must satisfy the fermionic pentagon equation, explicitly given in Eq.\eqref{fpentagon}.

The fermionic string-net model is defined on a trivalent lattice that is dual to a triangulation with a branching structure~\cite{gu2014lattice,gu2015classification,bhardwaj2016state}. This induces a direction on each edge of the trivalent graph such that every vertex is of fusion of splitting type. The fermionic string-net Hamiltonian is a sum of commuting projectors which implies that its ground state has an exact tensor network description. We consider the fPEPS tensor on a fusion vertex (the splitting case proceeds similarly)
\begin{equation}
\sum_{\substack{bcdhkl \\ \sigma \rho \nu \lambda}} [F^{bcd}_{k}]_{h\sigma\rho}^{l\nu\lambda} \pket{\nu_{l}^{cd}}\pbket{\lambda_{k}^{bl}} \pbbra{\rho^{hd}_{k} \sigma^{bc}_{h}}
\end{equation}
where we have used the shorthand notation ${\pket{\nu_{l}^{cd}}=\pket{\nu,c\, d,l}}$. Note only the degeneracy degree of freedom $\nu$ is fermionic. Double kets indicate physical indices of the fPEPS and single kets indicate virtual indices. This notation is used for consistency with the MPOs appearing in previous sections. 
To construct a globally valid fPEPS representation of the ground states of the  fermionic string-net models requires additional normalization factors be added to the above definition, see Ref.\cite{bultinck2017fermionic} for a full explanation. However, we will only use this tensor to demonstrate the simplest case of the pulling through equation, where these normalizations are irrelevant. 

The relevant fMPO tensor is given by
\begin{equation}
\sum_{\substack{abefkl \\ \lambda \mu \alpha \delta}} [F^{abl}_{e}]_{f\alpha \delta}^{k \lambda \mu} \ket{\lambda_{k}^{bl}}\bket{\mu_{e}^{ak}} \bra{\delta^{fl}_{e}} \bbra{\alpha^{ab}_{f}}
\end{equation}
where round kets indicate inner indices of the fMPO. In the above MPO the virtual index $e$ indicates which block we are in. Each individual block fMPO is injective, with inverse given by Eq.\eqref{finv}.

The fusion tensor for the single block fMPOs we have defined is
\begin{equation}
\sum_{\substack{ abcdef \\ \alpha \beta \sigma \psi }}
[F^{abc}_{g}]_{f\alpha \beta}^{h \sigma \psi} \bket{ \sigma^{bc}_{h} \psi^{ah}_{g}} \bbra{ \beta^{fc}_{g}} \sbra{\alpha^{ab}_{f}}
\end{equation}
where the double rounded bra indicates a fusion degeneracy index.

The \emph{zipper condition} for this fusion tensor reads
\begin{align}
&\sum [F^{abl}_{e}]_{f\alpha \delta}^{k \lambda \mu} \bket{\lambda_{k}^{bl} \mu_{e}^{ak}} \bbra{\delta^{fl}_{e}} \sbra{\alpha^{ab}_{f}} 
\nonumber \\
& \phantom{\sum\sum\sum\sum} \otimes [F^{fcd}_{e}]_{g \beta \gamma}^{l \nu \delta} \pbket{\nu_{l}^{cd}}\bket{\delta_{e}^{fl}} \pbbra{\gamma^{gd}_{e}} \bbra{ \beta^{fc}_{g}}
\nonumber \\
&= \sum
[F^{bcd}_{k}]_{h\sigma\rho}^{l\nu\lambda} \pbket{\nu_{l}^{cd}}\bket{\lambda_{k}^{bl}} \pbbra{\rho^{hd}_{k} } \bbra{\sigma^{bc}_{h}}
\otimes [F^{ahd}_{e}]_{g\psi\gamma}^{k\rho\mu} \pbket{\rho^{hd}_{k}}
 \nonumber \\
& \bket{\mu_{e}^{ak}}  \pbbra{\gamma^{gd}_{e}} \bbra{\psi^{ah}_{g}}
\otimes [F^{abc}_{g}]_{f\alpha \beta}^{h \sigma \psi}  \bket{ \sigma^{bc}_{h} \psi^{ah}_{g}} \bbra{ \beta^{fc}_{g}} \sbra{\alpha^{ab}_{f}}
\end{align}
where the summations are over all variables that appear. The equality follows from Eq.\eqref{fpentagon} after rearranging the fermionic modes and comparing coefficients.

The associator that changes the order of fusion is simply given by the input $F$-symbol for our chosen tensors. The $F$-move equation is
\begin{align}
&\sum [F^{abl}_{e}]_{f\alpha \delta}^{k \lambda \mu} \bket{\lambda_{k}^{bl} \mu_{e}^{ak}} \bbra{\delta^{fl}_{e}} \sbra{\alpha^{ab}_{f}}
\nonumber \\
& \phantom{\sum\sum\sum\sum}  \otimes[F^{fcd}_{e}]_{g \beta \gamma}^{l \nu \delta} \bket{\nu_{l}^{cd} \delta_{e}^{fl}} \bbra{\gamma^{gd}_{e}} \sbra{ \beta^{fc}_{g}}
\nonumber \\
&= \sum
[F^{bcd}_{k}]_{h\sigma\rho}^{l\nu\lambda} \bket{\nu_{l}^{cd} \lambda_{k}^{bl}} \bbra{\rho^{hd}_{k} } \sbra{\sigma^{bc}_{h}}
\otimes [F^{ahd}_{e}]_{g\psi\gamma}^{k\rho\mu} \bket{\rho^{hd}_{k} \mu_{e}^{ak}}
 \nonumber \\
& \phantom{\sum\sum\sum}
  \bbra{\gamma^{gd}_{e}} \sbra{\psi^{ah}_{g}}
\otimes [F^{abc}_{g}]_{f\alpha \beta}^{h \sigma \psi}  \sket{ \sigma^{bc}_{h} \psi^{ah}_{g}} \sbra{ \beta^{fc}_{g} \alpha^{ab}_{f}}
\end{align}
which follows from Eq.\eqref{fpentagon} after bringing the fermion modes into a common order. Note the $F$-symbol arising from the fusion of fMPOs in this equation ${[F^{abc}_{g}]_{f\alpha \beta}^{h \sigma \psi}  \sket{ \sigma^{bc}_{h} \psi^{ah}_{g}} \sbra{ \beta^{fc}_{g} \alpha^{ab}_{f}} }$ precisely matches the one used to build the fMPO tensors.
These $F$-symbols are consistent since the input symbol is a solution to Eq.\eqref{fpentagon}.

\begin{figure}[ht]
\center
\includegraphics[width=0.95\linewidth]{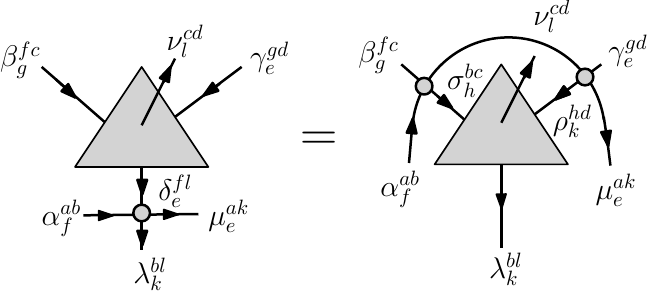} 
\caption{Graphical notation for the pulling through equation~\eqref{fspullthru}.}\label{onlyfig}
\end{figure}

Having defined the fPEPS and fMPO tensor one can now check that they satisfy the \emph{pulling through equation} shown in Fig.\ref{onlyfig}
\begin{align}
\label{fspullthru}
&\sum [F^{abl}_{e}]_{f\alpha \delta}^{k \lambda \mu} \pbket{\lambda_{k}^{bl}}\bket{\mu_{e}^{ak}} \pbbra{\delta^{fl}_{e}} \bbra{\alpha^{ab}_{f}}
\nonumber \\
& \phantom{\sum\sum\sum\sum} \otimes [F^{fcd}_{e}]_{g \beta \gamma}^{l \nu \delta} \pket{\nu_{l}^{cd}}\pbket{\delta_{e}^{fl}} \pbbra{\gamma^{gd}_{e} \beta^{fc}_{g}}
\nonumber \\
&= \sum
[F^{bcd}_{k}]_{h\sigma\rho}^{l\nu\lambda} \pket{\nu_{l}^{cd}}\pbket{\lambda_{k}^{bl}} \pbbra{\rho^{hd}_{k} \sigma^{bc}_{h}}
\otimes [F^{ahd}_{e}]_{g\psi\gamma}^{k\rho\mu} \pbket{\rho^{hd}_{k}}
 \nonumber \\
& \bket{\mu_{e}^{ak}}  \pbbra{\gamma^{gd}_{e}} \bbra{\psi^{ah}_{g}}
\otimes [F^{abc}_{g}]_{f\alpha \beta}^{h \sigma \psi} \pbket{ \sigma^{bc}_{h}}\bket{\psi^{ah}_{g}} \pbbra{ \beta^{fc}_{g}} \bbra{\alpha^{ab}_{f}}
\end{align}
which again follows from the fermionic pentagon equation~\eqref{fpentagon} for the $F$-symbols. This can be seen by bringing the fermion bases on either side into a common order and comparing the coefficients. We remark that it has subsequently been realized\cite{bultinck2017fermionic} that there is an extremely important subtlety for fMPO virtual symmetries of fPEPS. Due to the possibility of different boundary conditions, unlike the bosonic case, additional pulling through equations are required to ensure that an fPEPS has a given fMPO virtual symmetry. See Ref.\cite{bultinck2017fermionic} for a detailed explanation of the full set of fermionic pulling through equations. 

Just as in the bosonic case, the pulling through equation lies at the heart of topological order in tensor networks. For example, it allows one to construct multiple ground states on the torus by placing these fMPOs along non-trivial cycles on the virtual level of the tensor network. There is a nontrivial interplay between the possible closures and different spin structures on the resulting manifold, see Ref.\cite{bultinck2017fermionic}. 
The full set of pulling through conditions explained in Ref.\cite{bultinck2017fermionic} ensure that the fMPOs can move through the lattice and hence can be used to construct different locally indistinguishable ground states.

\section{ Conclusions}

In this paper we have initiated the study of two dimensional fermionic phases of matter with matrix product operators, extending beyond the bosonic case~\cite{schuch2010peps,buerschaper2014twisted,csahinouglu2014characterizing}.
The fundamental objects in our approach are fermionic MPOs that captures a reflection of topological order in the entanglement degrees of freedom of a PEPS.
Studying the algebra generated by these fMPOs led us to define local fusion tensors and associators that relate different orders of fusion.
We found that the associators must satisfy a consistency constraint known as the fermionic pentagon equation.

Hence fMPO algebras are classified by equivalence classes of solutions to the fermionic pentagon equation up to gauge transformation.
However, for topological phases we know that these equivalence classes do not directly correspond to the full classification of emergent excitations~\cite{bultinck2015anyons}. The physically relevant classification needs to take a further equivalence relation of the fMPOs known as Morita equivalence. This can be done following the approach of Ref.\cite{bultinck2015anyons}, where the superselection sectors are constructed directly in terms of irreducible central idempotents of a secondary fMPO algebra that depends on the fusion tensors. The explicit details of this construction will be reported in future work. Furthermore, by assuming the invertibility of individual fMPO tensors, the full interplay of the $F$-symbols with Majorana fMPOs is missed. This is explored more carefully in the follow up work Ref.\cite{bultinck2017fermionic}.

To classify symmetry respecting phases in the presence of a physical symmetry group $G$, corresponding to symmetry-enriched topological (SET) phases, the relevant Morita equivalence relation on the MPOs is weaker as it cannot alter the group structure~\cite{williamson2014matrix}. In the extreme case of SPT phases there is a single irreducible fMPO per group element. Hence the classification of fMPO group algebras is equivalent to the physical classification of SPT phases (with unitary on-site symmetry) which we found to be given by $H^1(G,\ZT),\,H^2(G,\ZT)$ and $\bar H^3(G,U(1))$ agreeing with known results~\cite{cheng2015towards,gu2014symmetry,kapustin2014fermionic}. We leave the study of SPTs with time reversal and lattice symmetries to future work.

A natural generalization of the framework presented here would be to consider $\mathbb{Z}_N$ parafermion models~\cite{fradkin1980disorder,fendley2012parafermionic,alicea2015topological}. This requires a $\mathbb{Z}_N$-grading rather than the $\ZT$-grading, as well as a modification of the fermionic swap rule. We remark that a recent work has appeared where parafermions are studied using MPS~\cite{xu2017matrix}, it would be interesting to extend this approach to MPO algebras.
\\ \\
\emph{Note added -}
Since the posting of this paper we have worked out the theory of fMPO symmetries of topological fPEPS in far greater detail. The results are given in Ref.\cite{bultinck2017fermionic} which supersede the preliminary discussion provided here. 
\\ \\
\emph{Acknowledgments -}
We acknowledge enlightening discussions with M. \surname{Mari\"en}.
D.W. acknowledges useful discussions with Z.-C. Gu while visiting Perimeter Institute as a graduate fellow. This work
was supported by the Austrian Science Fund (FWF) through grants ViCoM and FoQuS, and the EC through the grant QUTE. J.H. and F.V. acknowledge the support from the Research Foundation Flanders (FWO).

\bibliography{fermionmpos}

\appendix

\end{document}